\documentclass[aps,prc,reprint,superscriptaddress,nofootinbib]{revtex4-2}
\usepackage{amsmath}
\usepackage{amssymb}
\usepackage{graphicx}% Include figure files
\usepackage{dcolumn}% Align table columns on decimal point
\usepackage{bm}% bold math
\usepackage{braket}
\usepackage{mathrsfs}

\begin{document}

%\preprint{}

\title{
Generalizing the calculable ${\bm R}$-matrix theory and eigenvector continuation to the incoming wave boundary condition
}


%\author{}
%\thanks{}
%\affiliation{}

\author{Dong Bai}
\email{dbai@tongji.edu.cn}
%\email{dbai@alumni.itp.ac.cn}
\affiliation{School of Physics Science and Engineering, Tongji University, Shanghai 200092, China}%

\author{Zhongzhou Ren}
\email[Corresponding author: ]{zren@tongji.edu.cn}
\affiliation{School of Physics Science and Engineering, Tongji University, Shanghai 200092, China}%
\affiliation{Key Laboratory of Advanced Micro-Structure Materials, Ministry of Education, Shanghai 200092, China}

\begin{abstract}

The calculable $R$-matrix theory has been formulated successfully for regular boundary conditions with vanishing radial wave functions at the coordinate origins [P.\ Descouvemont and D.\ Baye, Rept.\ Prog.\ Phys.\ {\bf 73}, 036301 (2010)]. 
We generalize the calculable $R$-matrix theory to the incoming wave boundary condition (IWBC), which is widely used in theoretical studies of low-energy heavy-ion fusion reactions to simulate the strong absorption of incoming flux inside the Coulomb barriers. 
The generalized calculable $R$-matrix theory also provides a natural starting point to extend eigenvector continuation (EC) [D.~Frame \emph{et al.}, Phys.\ Rev.\ Lett.\ {\bf 121}, 032501 (2018)]
to fusion observables.
The ${}^{14}\text{N}+{}^{12}\text{C}$ fusion reaction is taken as an example to validate these new theoretical tools. 
Both local and nonlocal potentials are considered in numerical calculations.
Our generalizations of the calculable $R$-matrix theory and EC are found to work well for IWBC.

%Both of them are found to work well for fusion reactions.

\end{abstract}

%\keywords{}

\maketitle

\section{Introduction}

The calculable $R$-matrix theory provides a powerful framework to solve the Schr\"odinger equations in regular boundary conditions \cite{Descouvemont:2010cx,Lane:1948zh,Burke:2011,Thompson:2009}, 
where radial wave functions vanish at the coordinate origins
and can take different asymptotic forms in different problems.
The calculable $R$-matrix theory divides the configuration space into internal and external regions by the channel radius. Compared to the internal region, physics in the external region gets simplified thanks to the negligibility of short-range interactions therein, and external wave functions are known explicitly up to a few coefficients.
The Bloch operator is often adopted to match internal and external wave functions continuously at the channel radius \cite{Bloch:1957}. 
The resultant Bloch-Schr\"odinger equations are then solved by, e.g., the variational method.
In nuclear physics, the calculable $R$-matrix theory has been used successfully to study bound states, resonant states, elastic/inelastic scatterings, transfer reactions, breakup reactions, and fusion reactions.
See Refs.~\cite{Bai:2020hmz,Assuncao:2013uma,Lei:2020zif,Descouvemont:2015xoa,Bai:2020,Thompson:2019uok,Shubhchintak:2019pmd,Li:2018zdo} for some recent works.

The incoming wave boundary condition (IWBC) plays a fundamental role in modern theoretical studies of low-energy heavy-ion fusion reactions \cite{Hagino:1999xb,Hagino:2012cu,Back:2014ypa}. 
Consider the Schr\"odinger equation
\begin{align}
&\left[-\frac{1}{2\mu}\frac{\mathrm{d}^2}{\mathrm{d}r^2}+\frac{L(L+1)}{2\mu r^2}+V_C(r)-E\right]\chi_L(r)\nonumber\\
=-&\begin{cases}
\ V_N(r)\,\chi_L(r),\\[2ex]
\ \int\!\mathrm{d}r'\,W^{(L)}_N(r,r')\,\chi_L(r'),
\end{cases}
\label{SE}
\end{align}
with $L$ being the orbital angular momentum, $\mu$ being the two-body reduced mass, $E>0$ being the reaction energy in the center-of-mass (CM) frame, $V_C(r)=Z_PZ_Te^2/r$ being the Coulomb potential, $V_N(r)$ and $W^{(L)}_N(r,r')$ being the local and nonlocal nuclear potentials, and $\chi_{L}(r)$ being the radial wave function. 
IWBC is given by
\begin{align}
\!\!\!\!\chi_L(r) & \sim T_L(E)\exp\!\left[-i\!\int_{r_\text{abs}}^r\!\!\!\!\!\!\mathrm{d}r'\,k_L(r')\right], \ 0\leq r\leq r_\text{abs},\label{IWBC1}\\
&=H_L^{(-)}(\eta,kr)-S_L(E)H_L^{(+)}(\eta,kr),\ r\geq r_\text{c}.\label{IWBC2}
\end{align}
%\begin{align}
%\chi_L(r) & =0, && r= 0,\label{OBC1}\\
%&=H_L^{(-)}(\eta,kr)+S_L(E)H_L^{(+)}(\eta,kr), && r\geq r_\text{c}.\label{OBC2}
%\end{align}
Here, $r_\text{abs}$ is the absorption radius inside the Coulomb barrier, $r_\text{c}$ is the channel radius chosen to be so large that the nuclear interaction and antisymmetrization between the target and the projectile become negligible in the external region, $k=\sqrt{2\mu E}$ is the relative momentum at $r\to\infty$, $k_L(r)$
%$=\sqrt{2\mu\left[E-\frac{L(L+1)}{2\mu r^2}-V_N(r)-V_C(r)\right]}$ 
is the relative momentum at $r\leq r_\text{abs}$, $\eta$ is the Sommerfeld parameter, $H_L^{(\mp)}(\eta,kr)$ are the incoming/outgoing Coulomb-Hankel functions, and $S_L(E)$ and $T_L(E)$ are the $S$- and transmission matrix elements.
The relative momentum $k_L(r)$ could be estimated by $k_L(r)\!\!=\!\!\sqrt{2\mu\!\left[E-\frac{L(L+1)}{2\mu r^2}-V_N(r)-V_C(r)\right]}$ for the local potential
and $k_L(r)\!=\!\!\sqrt{2\mu\!\left[E-\frac{L(L+1)}{2\mu r^2}-W^\text{LE}_N(r)-V_C(r)\right]}$ for the nonlocal potential,
with $W_N^\text{LE}$ being some local equivalence of the nonlocal potential (see Section \ref{NR}).
IWBC is different from regular boundary conditions which impose $\chi_L(r)=0$ at $r=0$. 
It is widely used in nuclear fusion problems to simulate the strong absorption of incoming flux inside the Coulomb barriers and has become one of the standard ansatzes to calculate fusion observables. Compared with the regular-boundary-condition approach to nuclear fusion reactions, IWBC is often regarded as more predictive in the sense that no extra imaginary optical potential is needed.

% and a large number of equidistant mesh points are needed to achieve numerical convergence. 
In this work, the generalized calculable $R$-matrix theory is proposed to solve the Schr\"odinger equations in IWBC.
%generalize calculable $R$-matrix theory from regular boundary conditions to IWBC, which could be a helpful complement to existing methods.
%As shown later on, the calculable $R$-matrix theory generally has the advantage to need fewer mesh points than the modified Numerov method.
Besides the academic interest to enlarge the applicable scope of the calculable $R$-matrix theory,
it provides a unified framework for both local and nonlocal potentials in nuclear fusion studies.
Nonlocal potentials have important applications in nuclear physics dating back to Perey and Buck in the 1960s \cite{Perey:1962} and get revived in recent years \cite{Titus:2016gvp,Waldecker:2016opm,Lovell:2017rzm,Tian:2018xby,Jaghoub:2018vrw,Velasquez:2019aau,Blanchon:2020ioa,Quinonez:2020dzj,Rotureau:2016jpf}.
Nucleus-nucleus nuclear interactions are intrinsically nonlocal in coordinate space thanks to the antisymmetrization effect 
and model-space truncations \cite{CandidoRibeiro:1997gp,Chamon:2002mx}.
Moreover, modern realistic interactions of nucleons based on chiral effective field theory 
are generally nonlocal \cite{Epelbaum:2008ga,Machleidt:2011zz,Hammer:2019poc},
which might also contribute to nonlocality in nucleus-nucleus potentials.
Nonlocal potentials have been adopted by some authors to study low-energy heavy-ion fusion reactions.
% \cite{Chamon:2007bb,Sastry:1997zz,Galetti:1994gn,Dutt:1996zz,Galetti:1994jm,Canto:2013}.
%In these studies, the absorption of incoming waves inside Coulomb barriers is, however, simulated by imaginary parts of optical potentials instead of IWBC. 
%It is important to study impacts of nonlocal potentials on low-energy heavy-ion fusion reactions \cite{Chamon:2007bb,Sastry:1997zz,Galetti:1994gn,Dutt:1996zz,Galetti:1994jm,Canto:2013}.
Refs.~\cite{Sastry:1997zz,Galetti:1994gn,Dutt:1996zz,Galetti:1994jm} study fusion problems with nonlocal potentials in the framework of the WKB approximation.
Refs.~\cite{Chamon:2007bb,Canto:2013} solve the Schr\"odinger equations with nonlocal complex optical potentials in regular boundary conditions.
As far as we know, there is few publication on how to solve the Schr\"odinger equations with nonlocal potentials and IWBC exactly.
The popular implementation of IWBC in the CCFULL code \cite{Hagino:1999xb} is based on the modified Numerov method \cite{Melkanoff:1966} and deals with local potentials only.
It is not easy to extend the method to nonlocal potentials.
Our generalized $R$-matrix theory fills this gap and gives a valuable opportunity to study the impacts of nonlocal potentials in nuclear fusion reactions.

The generalized calculable $R$-matrix theory also provides a natural starting point to extend eigenvector continuation (EC) \cite{Frame:2017fah}, a variational emulator for bound-state observables, to fusion observables.
EC is characterized by choosing basis functions from the Hamiltonian eigenstates at selected training points in the parameter space.
It has been shown to be a reliable and efficient tool in uncertainty quantification and global sensitivity analysis \cite{Frame:2017fah,Frame:2019jsw,Konig:2019adq,Ekstrom:2019lss,Demol:2019yjt,Sarkar:2020mad}, where model evaluations, sometimes computationally expensive, have to be carried out for a large number of times at different points in the parameter space.
An on-going direction of EC is to extend the method to reaction observables.
Recently, some progress has been made for scattering observables via the Kohn variational principle \cite{Furnstahl:2020abp}.
Generally speaking, the calculable R-matrix theory and its generalizations provide a comprehensive roadmap to extend EC from bound-state observables to other structural and reaction observables, not limited to fusion observables.
Here, we just focus on fusion observables for concreteness.

The rest parts are organized as follows. In Section \ref{TF}, our generalizations of the calculable $R$-matrix theory and EC are presented.
In Section \ref{NR}, numerical reliability of these methods is examined in detail by studying the ${}^{14}\text{N}+{}^{12}\text{C}$ fusion reaction. Conclusions are given in Section \ref{Concl}. The natural units $\hbar=c=1$ are adopted in this work.

\begin{widetext}

\section{Theoretical Formalism}
\label{TF}

\subsection{The Generalized Calculable $\bm R$-Matrix Theory}

%\begin{widetext}

\begin{figure*}

  \includegraphics[width=0.8\linewidth]{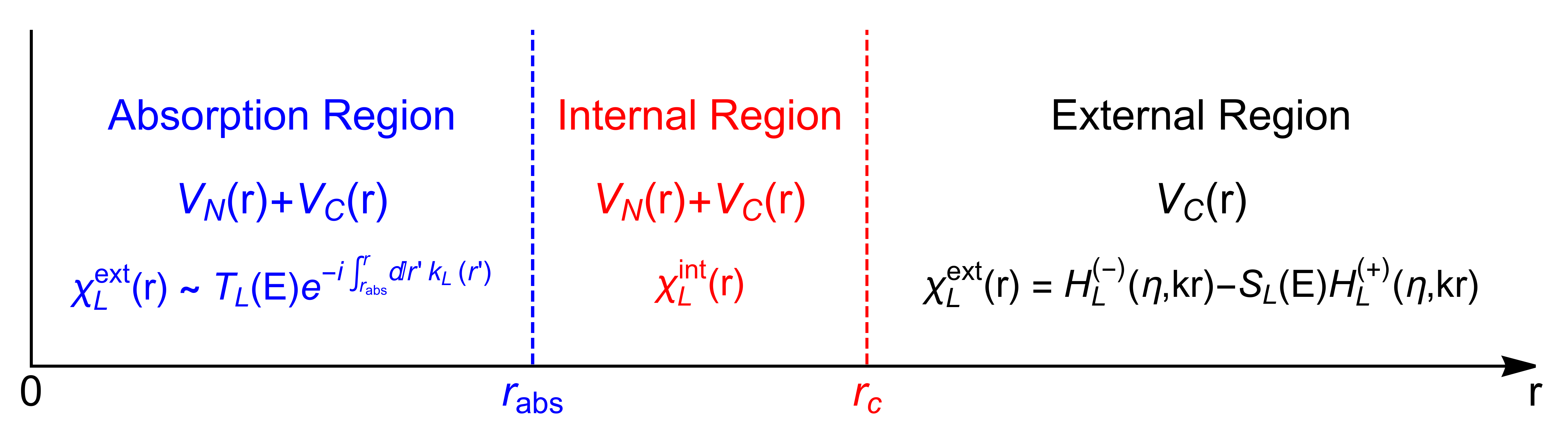}
  
  \caption{The trisection of the configuration space in the generalized calculable $R$-matrix theory for IWBC.
  In Absorption and Internal Regions, both the nuclear and Coulomb potentials are sizable, 
  while in External Region, only the Coulomb potential is sizable.
  The nuclear potential is taken to be the local potential $V_N(r)$ for simplicity. 
The trisection of the configuration space works similarly for nonlocal potentials.}
  \label{Illus}
  
  \vskip -0.5cm
  
\end{figure*}

%\end{widetext}

We divide the configuration space into three parts, ``Absorption Region" $[0,r_\text{abs}]$, ``Internal Region" $[r_\text{abs},r_\text{c}]$, and ``External Region'' $[r_\text{c},\infty)$. See Fig.~\ref{Illus} for an illustration. In Absorption and Internal Regions, both the nuclear potential $V_N(r)$ and the Coulomb potential $V_C(r)$ are sizable, while in External Region  the nuclear potential becomes negligible and only the Coulomb potential predominates. External wave functions in Absorption and External Regions are given by IWBC in Eqs.~\eqref{IWBC1}-\eqref{IWBC2}, which should be matched continuously to internal wave functions in Internal Region.

%\subsubsection{Single-Channel Formalism}

In the calculable $R$-matrix theory, the Schr\"odinger equation in Eq.~\eqref{SE} is promoted to the Bloch-Schr\"odinger equation
\begin{align}
%\!\!
\left[-\frac{1}{2\mu}\frac{\mathrm{d}^2}{\mathrm{d}r^2}+\frac{L(L+1)}{2\mu r^2}+\mathcal{L}(r_\text{c})-\mathcal{L}(r_\text{abs})+V_C(r)-E\right]\chi_L^\text{int}(r)%\nonumber\\
%&\qquad\qquad\qquad\qquad\qquad\qquad\qquad\qquad\qquad\qquad\qquad\qquad
=&\left[\mathcal{L}(r_\text{c})-\mathcal{L}(r_\text{abs})\right]\chi_L^\text{ext}(r)\nonumber
\end{align}
\begin{align}
%\qquad\qquad\qquad\qquad\qquad\qquad\qquad\qquad\qquad\qquad\qquad\qquad\qquad\qquad\qquad
-&\begin{cases}
\ V_N(r)\,\chi^\text{int}_L(r),\\[2ex]
\ \int_{r_\text{abs}}^{r_\text{c}}\!\mathrm{d}r'\,W^{(L)}_N(r,r')\,\chi^\text{int}_L(r'),
\end{cases}
\label{BSE}
\end{align}
with $\mathcal{L}(R)=\frac{1}{2\mu}\delta(r-R)\frac{\mathrm{d}}{\mathrm{d}r}$ being the Bloch operator. 
It is easy to verify that $-\frac{1}{2\mu}\frac{\mathrm{d}^2}{\mathrm{d}r^2}+\mathcal{L}(r_\text{c})-\mathcal{L}(r_\text{abs})$ is hermitian in Internal Region $[r_\text{abs},r_\text{c}]$.
Besides, the continuity of wave functions imposes
$\chi_L^\text{int}(r_\text{abs})=\chi_L^\text{ext}(r_\text{abs})$ and $\chi_L^\text{int}(r_\text{c})=\chi_L^\text{ext}(r_\text{c})$.
The Bloch-Schr\"odinger equation in Eq.~\eqref{BSE} could be solved by the variational method with $\chi_L^\text{int}(r)=\sum_{n=1}^{N}c_n(E)\varphi_n(r)$. 
%In this work, two basis functions are used explicitly, i.e., Gaussian and Lagrange functions.
The transmission matrix element $T_L(E)$ and the $S$-matrix element $S_L(E)$ are obtained by solving the linear equations
\begin{align}
&\sum_{n=1}^{N}C_{mn}(E)\,c_{n}(E)=%
%\left(\varphi_m\Big|\mathcal{L}(r_\text{c})-\mathcal{L}(r_\text{abs})\Big|\chi_L^{\text{ext}}\right),
\frac{1}{2\mu}\left\{k\varphi_m(r_\text{c})\!\left[H_L^{{(-)}'}\!(\eta,kr_\text{c})-S_L(E)H_L^{{(+)}'}\!(\eta,kr_\text{c})\right]\!+i T_L(E)k_L(r_\text{abs})\varphi_m(r_\text{abs})\right\},
\label{ME1}\\
&\sum_{n=1}^{N}\varphi_n(r_\text{abs})\,c_{n}(E)=T_L(E),\label{ME2}\\
&\sum_{n=1}^{N}\varphi_n(r_\text{c})\,c_{n}(E)=H_L^{(-)}(\eta,kr_\text{c})-S_L(E)H_L^{(+)}(\eta,kr_\text{c}),\label{ME3}
\end{align}
with 
\begin{align}
&C_{mn}(E)
=\left(\varphi_m\left|-\frac{1}{2\mu}\frac{\mathrm{d}^2}{\mathrm{d}r^2}+\frac{L(L+1)}{2\mu r^2}+\mathcal{L}(r_\text{c})-\mathcal{L}(r_\text{abs})+V_C(r)-E\right|\varphi_n\right)+
\begin{cases}
&\!\!\!\! \left(\varphi_m\left|V_N(r)\right|\varphi_n\right),\\[2ex]
&\!\!\!\! \left(\varphi_m\left|W_N^{(L)}(r,r')\right|\varphi_n\right).
\end{cases}
\end{align}
Here, the round bracket denotes the inner product over Internal Region $[r_\text{abs},r_\text{c}]$, i.e.,
%\begin{align}
$\left(\phi\left|\mathcal{O}\right|\psi\right)=\int_{r_\text{abs}}^{r_\text{c}}\!\mathrm{d}r\,\phi(r)\mathcal{O}(r)\psi(r)$.
In the calculable $R$-matrix theory, it is often convenient to use the Lagrange functions $\{\mathbb{L}^{\!\!N}_i(r)\}$ \cite{Baye:2015} as variational basis functions 
(see Appendix \ref{LF}).
Noticeably, the Lagrange functions used here are a bit different from the ordinary Lagrange functions $\mathbb{L}^{\!\!N}_i(r)\propto r-r_\text{abs}$ defined in the same interval $[r_\text{abs},r_c]$. The latter always satisfy 
$\mathbb{L}^{\!\!N}_i(r)=(-1)^{N+i}\frac{r-r_\text{abs}}{\Delta r x_i}\sqrt{x_i(1-x_i)\Delta r}\frac{\mathbb{P}_N[(2r-r_\text{abs}-r_\text{c})/\Delta r]}{r-x_i\Delta r-r_\text{abs}}
$,
with $\mathbb{P}_N(x)$ being the Legendre polynomial of order $N$ and $\Delta r=r_\text{c}-r_\text{abs}$.
%$\mathbb{L}^{\!\!N}_i(r_\text{abs})=0$.
This means that any finite combination of the ordinary Lagrange functions becomes exactly zero at the absorption radius and violates IWBC by construction.
As a result, the matching between the wave functions in Absorption and Internal Regions cannot be handled properly with the ordinary Lagrange functions.
In comparison, the Lagrange functions defined in Appendix \ref{LF} give nonzero values at the absorption radius and thus are better suited to matching the wave functions in Absorption and Internal Regions.
The numerical calculations in Section \ref{NR} show explicitly that 
the Lagrange functions defined in Appendix \ref{LF} are suitable for our purpose.
There are other choices for the variational basis functions.
For example, we have checked that the Gaussian basis functions $\varphi_n(r)=\exp(-\nu_nr^2)$ could be used as the basis functions as well.
However, the corresponding matrix elements have to be calculated by numerical integration, except in some specific cases.
This makes the Gaussian basis functions numerically less friendly than the Lagrange functions.
As shown in Appendix \ref{LF}, all the relevant matrix elements could be calculated analytically with the Lagrange functions,
which is an important advantage in numerical calculations.
The above theoretical formalism might turn out to be a bit similar to the $R$-matrix propagation method \cite{Light:1976}, but it has different theoretical motivations and application scenarios.
With the transmission matrix element $T_L(E)$, the fusion cross section is given by
\begin{align}
&\sigma_\text{fus}(E)=\sum_{L=0}^{L_\text{max}}\sigma_L(E)=\frac{\pi}{k^2}\sum_{L=0}^{L_\text{max}}(2L+1)P_L(E),\\
&P_L(E)=\frac{k_L(r_\text{abs})}{k}\left|T_L(E)\right|^2,
\end{align}
with $L_\text{max}$ being the maximal partial-wave angular momentum taken into account.
The resultant uncertainties are referred to as truncation errors.

\end{widetext}

\begin{figure*}

  \includegraphics[width=0.45\linewidth]{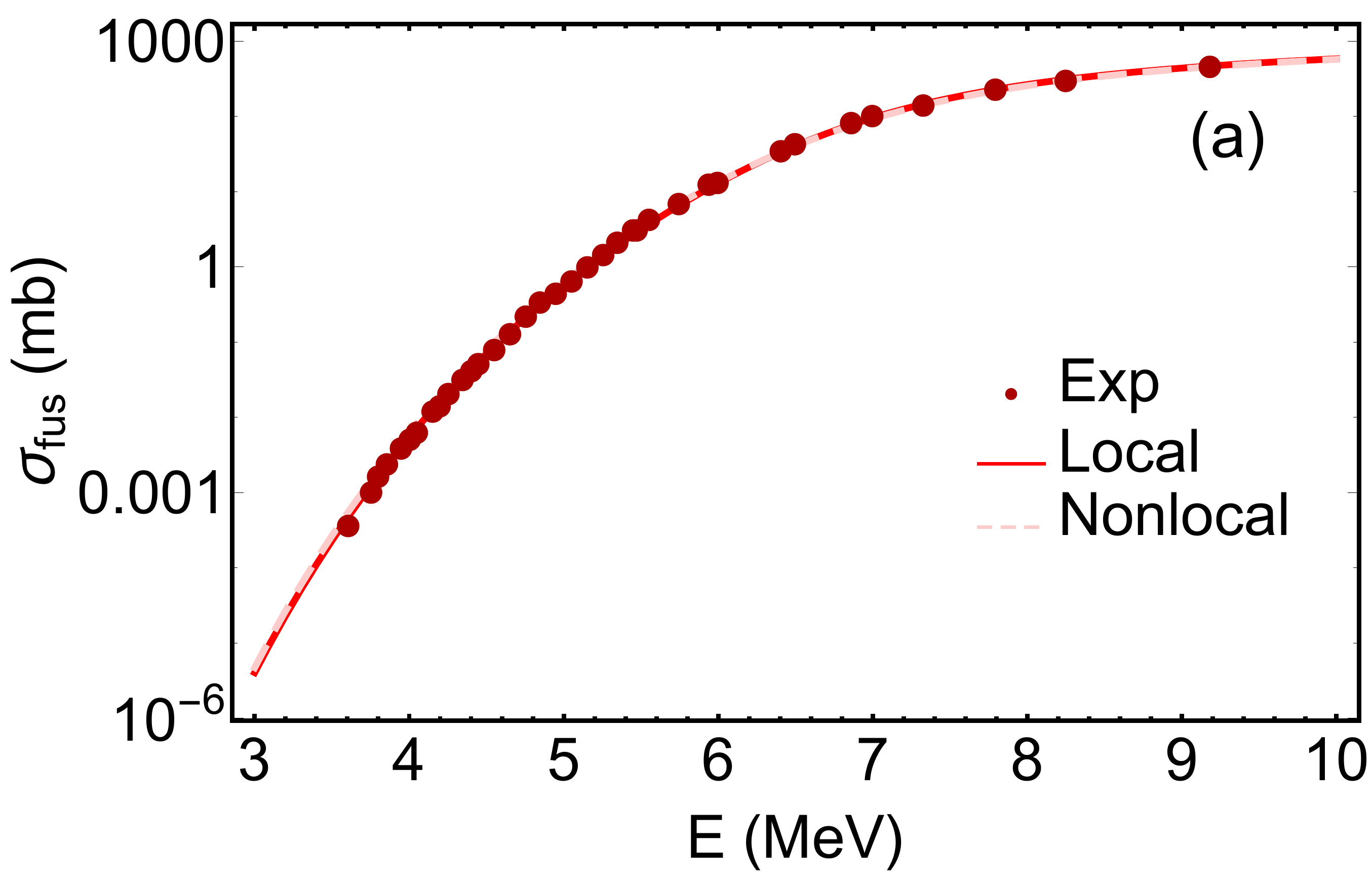}\\
%    \includegraphics[width=0.45\linewidth]{NR_R-Matrix.pdf}
%  \vskip 0.2cm
  \includegraphics[width=0.45\linewidth]{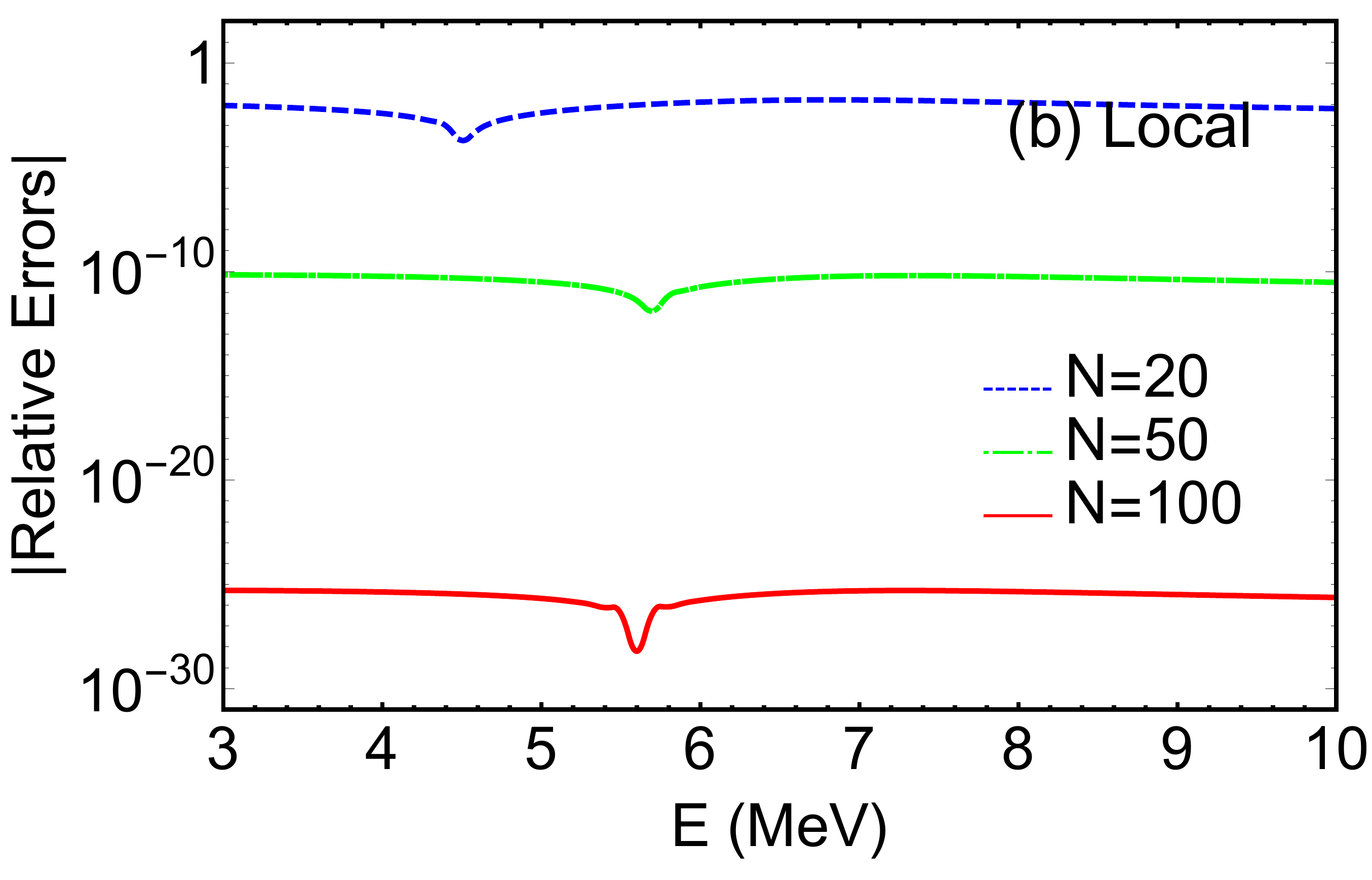}
    \includegraphics[width=0.45\linewidth]{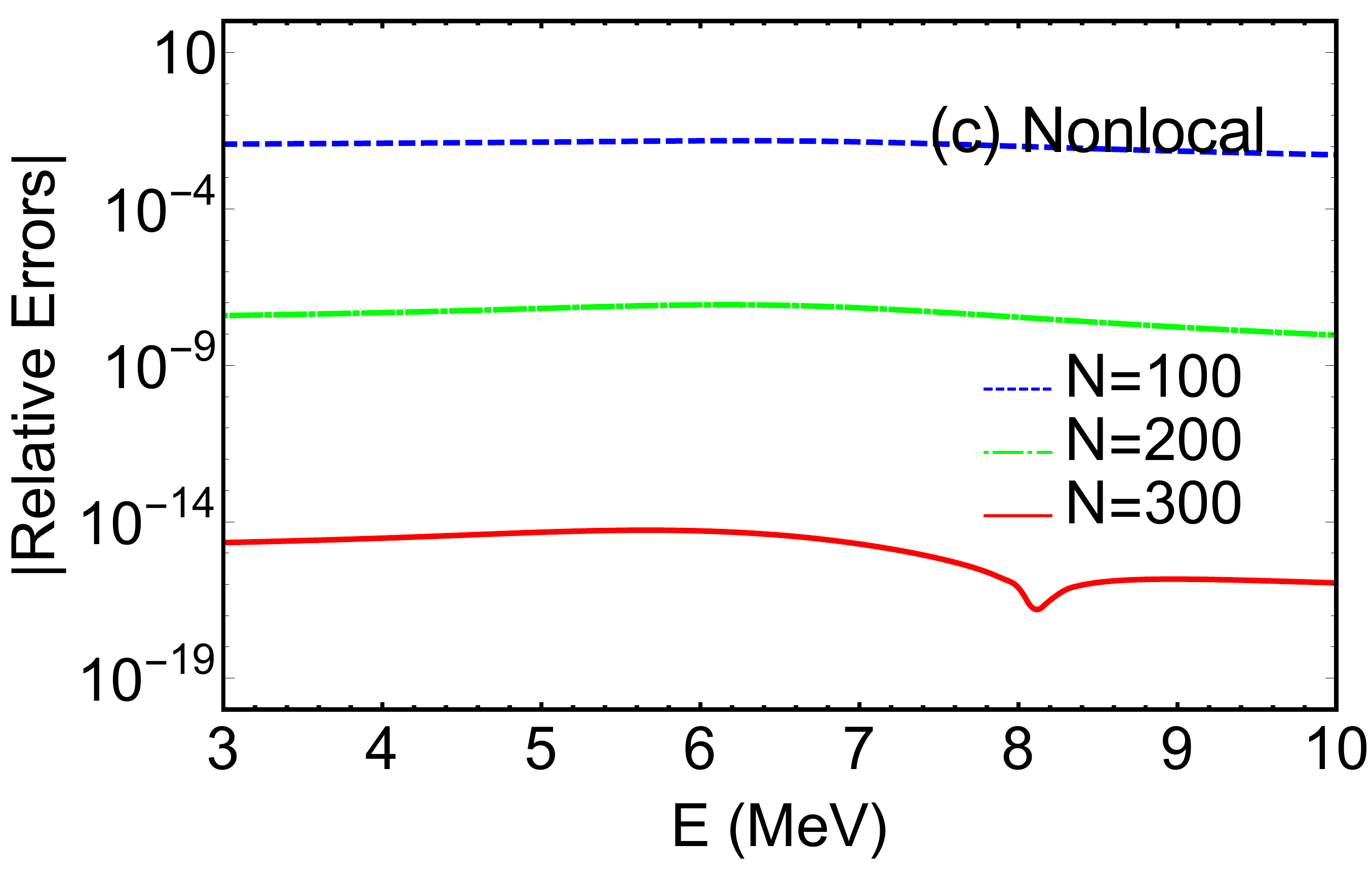}
    
    \vskip -0.3cm
  
  \caption{(a) Theoretical fusion cross sections of the ${}^{14}\text{N}+{}^{12}\text{C}$ fusion reaction at different reaction energies in the CM frame given by the generalized calculable $R$-matrix theory for the local potential in Eq.~\eqref{WS} (solid red line) and the nonlocal potential in Eq.~\eqref{PB} (dashed pink line).
   The parameters are given by 
   $V_0=60 \text{ MeV}$, $R_0=1.206(A_T^{1/3}+A_P^{1/3}) \text{ fm}$, $a=0.5 \text{ fm}$
   in Eq.~\eqref{WSP}. 
$N\geq20$ and $N\geq100$ Lagrange functions are used in the numerical calculations for local and nonlocal potentials, respectively.  
The dark red points are the experimental data from Ref.~\cite{Switkowski:1977wsf}. 
(b) Absolute values of the relative errors of the fusion cross sections given by the generalized calculable $R$-matrix theory with the local potential.
% $N$ is the number of Lagrange functions used in numerical calculations.
The benchmark values to calculate relative errors are given by the generalized calculable $R$-matrix theory with $N=200$ Lagrange functions.
(c) Absolute values of the relative errors of the fusion cross sections given by the generalized calculable $R$-matrix theory with the nonlocal potential.
$N=400$ Lagrange functions are used to calculate the benchmark values.
}
  \label{RM}
  
    \vskip -0.5cm
  
\end{figure*}

%\begin{figure}
%
%    \includegraphics[width=\linewidth]{EC-rmax-dependence.pdf}
%    
%      \caption{Theoretical results for the local potential of the Woods-Saxon form in Eqs.~\eqref{WS} and \eqref{WSP} at different channel radii: (a) fusion cross sections in the generalized $R$-matrix theory with $N=20, 50, 100, 200, 400, 800$ Lagrange mesh points and (b) absolute values of relative errors of fusion cross sections in the generalized EC with the same $N_\text{EC}=6, 10$ basis functions used in Fig.~\ref{EC}.}
%
%\label{SpeedUp}
%
%\end{figure}

\subsection{The Generalized Eigenvector Continuation}

The Bloch-Schr\"odinger equation in Eq.~\eqref{BSE} could be rewritten schematically as
\begin{align}
\bm{\mathcal{H}}_E(\bm{\alpha})\chi_L^\text{int}(\bm{\alpha})=\bm{\mathcal{L}}\chi_L^\text{ext}(\bm{\alpha}).
\end{align}
Here, $\bm{\mathcal{H}}_E(\bm{\alpha})$ and $\bm{\mathcal{L}}$ stand for the operators acting on the internal and external wave functions, with $\bm{\alpha}$ being the model parameters. 
The subscript ``$E$'' stresses that the operator $\bm{\mathcal{H}}_E(\bm{\alpha})$ depends on the reaction energy $E$. 
Let $\{\chi_L^\text{int}(\bm{\alpha}^\text{tr}_i)\}$ be the exact internal wave functions of the Bloch-Schr\"odinger equations at the training points $\{\bm{\alpha}^\text{tr}_i\}$ in the parameter space.
Following the philosophy of EC, we construct the variational emulator $\chi_L^\text{int}(\bm{\alpha}^\text{te}_\odot)$ for the test point $\bm{\alpha}^\text{te}_\odot$
\begin{align}
\chi_L^\text{int}(\bm{\alpha}^\text{te}_\odot)=\sum_{i=1}^{N_\text{EC}}c_i\,\chi_L^\text{int}(\bm{\alpha}^\text{tr}_i).
\label{GEC}
\end{align}
The coefficients $\{c_i\}$ and fusion observables are obtained by solving Eqs.~\eqref{ME1}-\eqref{ME3} with the EC basis functions $\{\chi_L^\text{int}(\bm{\alpha}^\text{tr}_i)\}$.

\begin{figure*}

  \includegraphics[width=0.46125\linewidth]{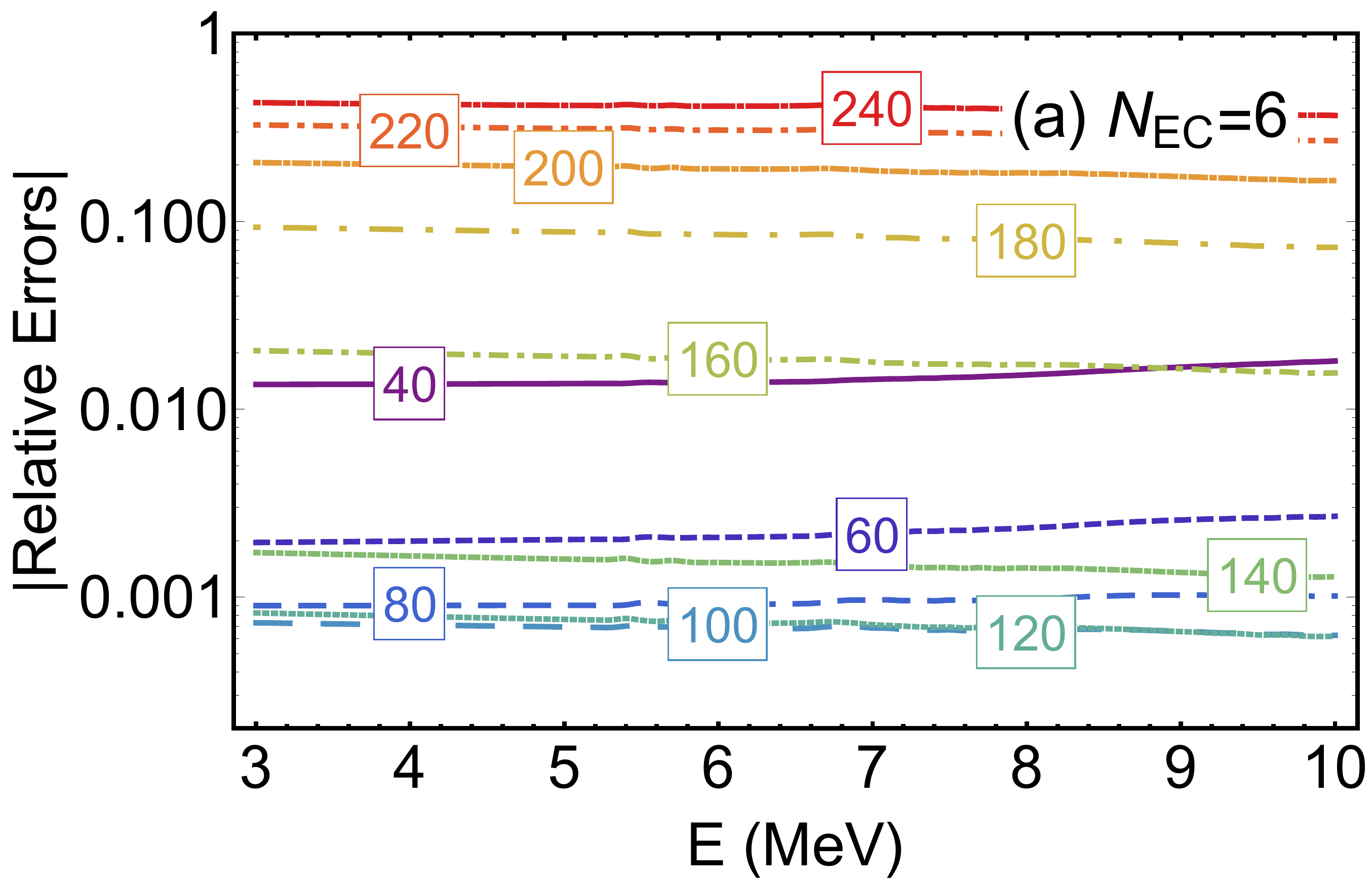}
%    \includegraphics[width=0.45\linewidth]{Relative_Error_EC_NEC_6.pdf}
%  \vskip 0.25cm
%  \includegraphics[width=0.45\linewidth]{Relative_Error_EC_NEC_10.pdf}
  \includegraphics[width=0.45\linewidth]{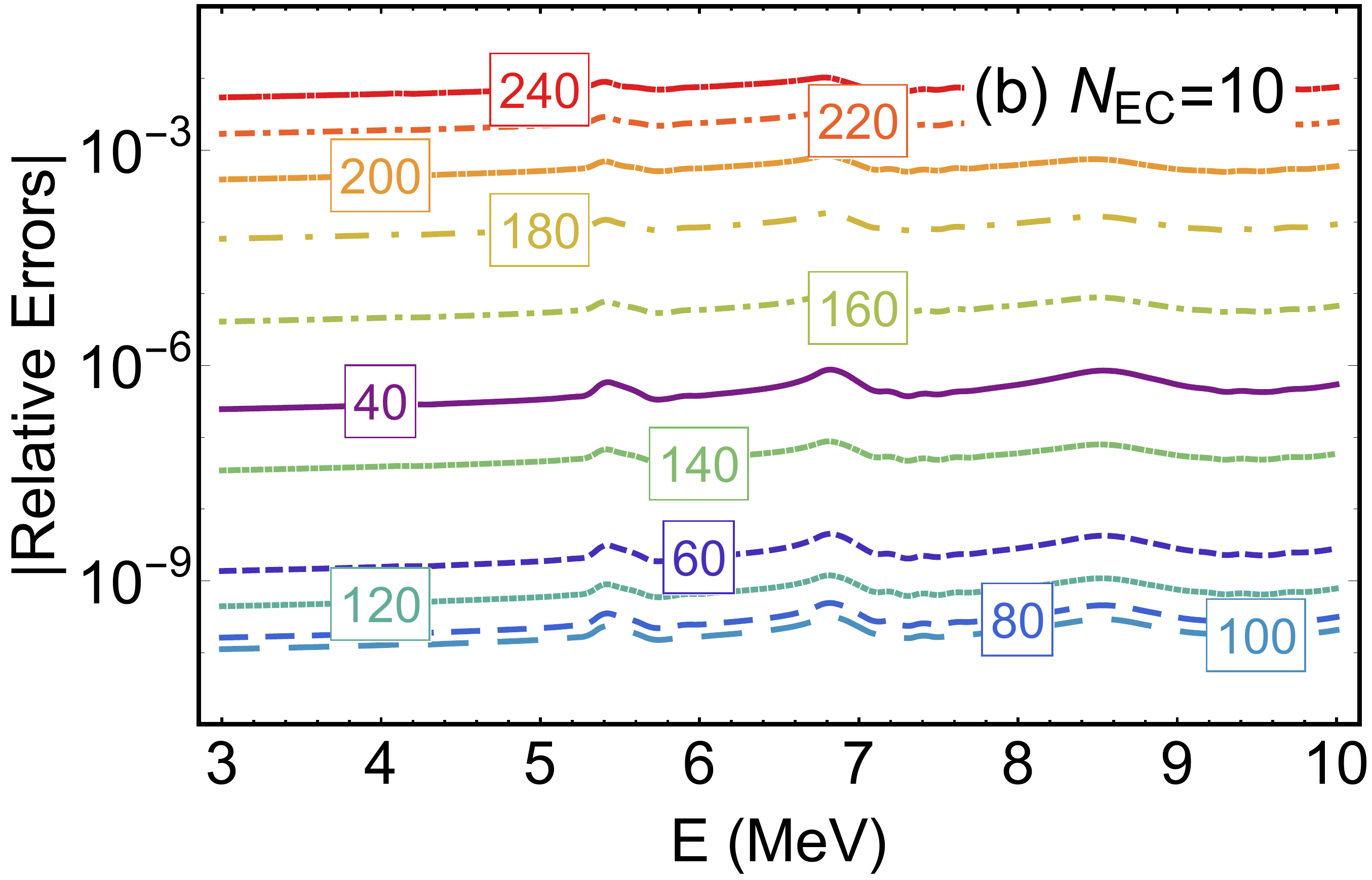}
  
  \caption{Absolute values of the relative errors of the fusion cross sections in the ${}^{14}\text{N}+{}^{12}\text{C}$ fusion reaction given by the generalized EC. The EC basis functions are taken to be the exact internal wave functions at (a) $N_\text{EC}=6$ training points $V_0=45, 65, \cdots, 125, 145$ MeV and (b) $N_\text{EC}=10$ training points $V_0=45, 55, \cdots, 125, 135$ MeV.
  The test points are taken to be $V_0=40, 60, \cdots, 220, 240$ MeV.
  $\boxed{\mbox{X}}$ is the abbreviation of $V_0=\text{X}$ MeV. }
  
  \label{EC}
        
\end{figure*}

\begin{figure*}

  \includegraphics[width=0.49125\linewidth]{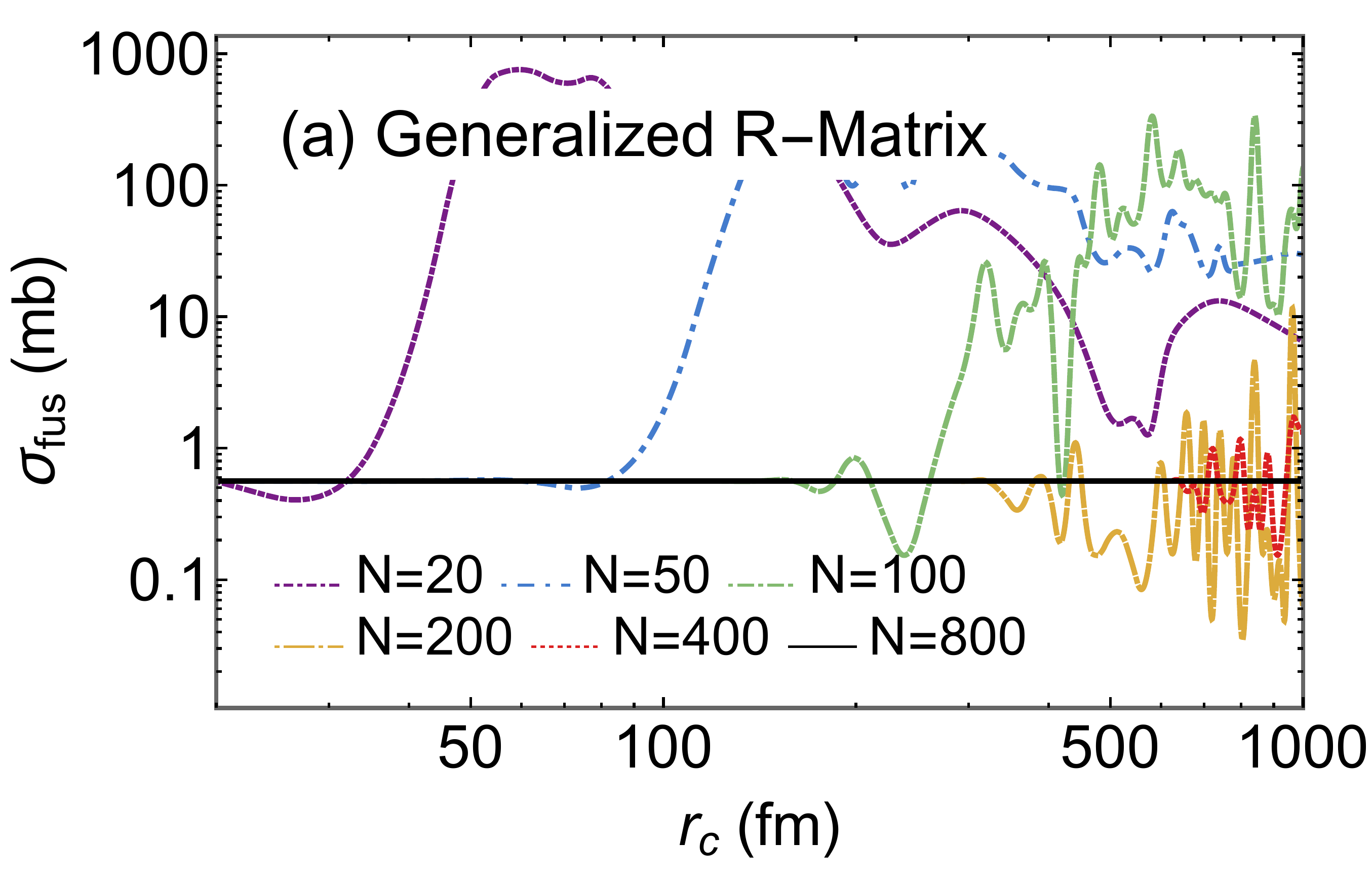}
%    \includegraphics[width=0.45\linewidth]{R-Matrix-rmax-dependence.pdf}
%\vskip 0.25cm
  \includegraphics[width=0.48\linewidth]{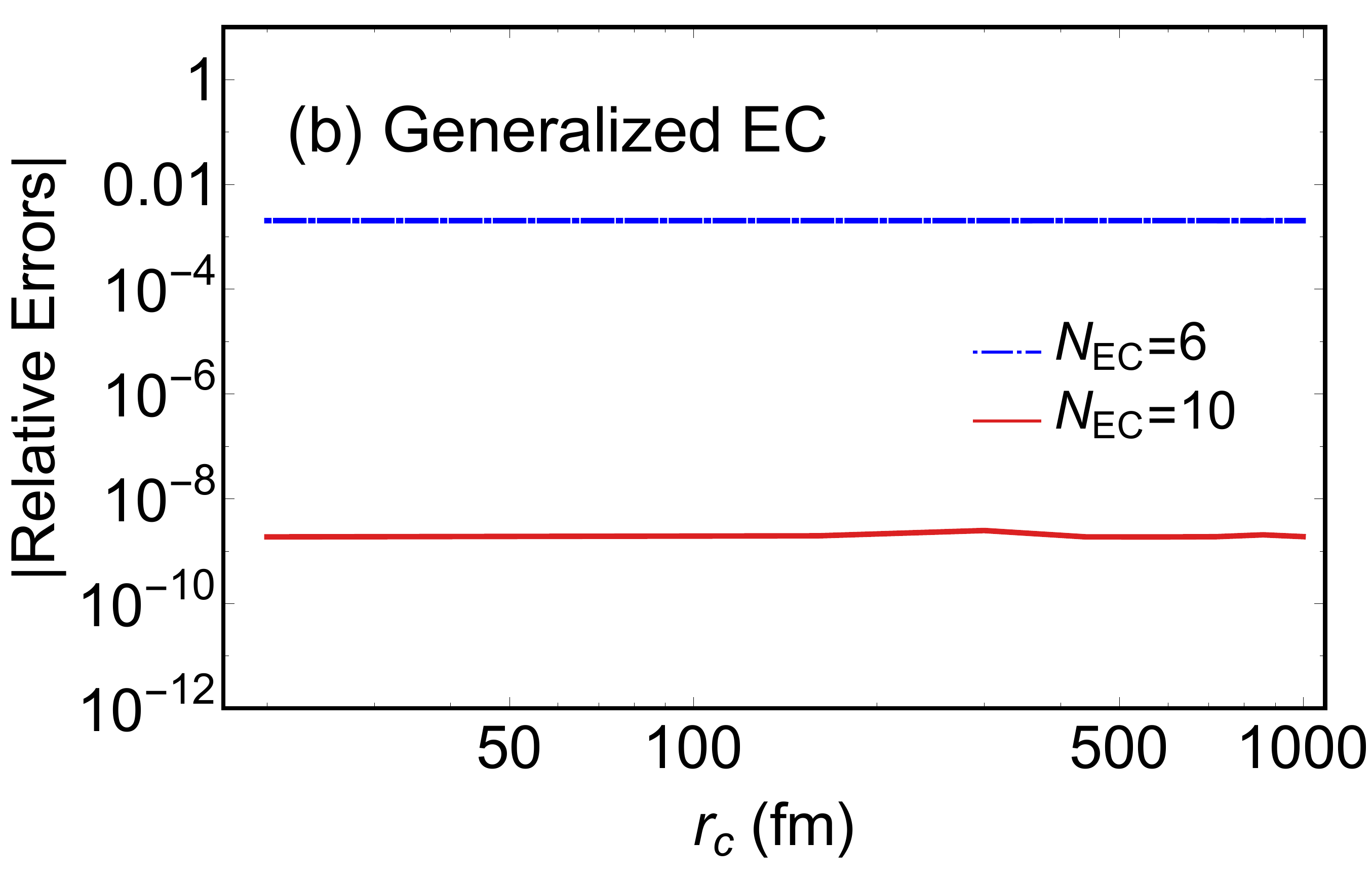}

  \caption{Theoretical results for the local potential in Eqs.~\eqref{WS} and \eqref{WSP} at different channel radii between 20 fm and 1000 fm: (a) the fusion cross sections given by the generalized $R$-matrix theory with $N=20, 50, 100, 200, 400, 800$ Lagrange functions and (b) absolute values of the relative errors of fusion cross sections given by the generalized EC with $N_\text{EC}=6, 10$ EC basis functions at the same training points as Fig.~\ref{EC}.}
  
  \label{rmax}
  
  \vskip -0.5cm
  
\end{figure*}

\section{Numerical Results}
\label{NR}

We take the ${}^{14}\text{N}+{}^{12}\text{C}$ fusion reaction as a proof-of-concept example to test our generalizations of the calculable $R$-matrix theory and EC. 
Both local and nonlocal potentials are used to describe nuclear interactions between ${}^{14}\text{N}$ and ${}^{12}\text{C}$.
The local nuclear potential $V_N(r)$ is taken to be the Woods-Saxon form
\begin{align}
V_N(r)=-\frac{V_0}{1+\exp[(r-R_0)/a]},
\label{WS}
\end{align}
while the nonlocal nuclear potential $W_N(\bm{r},\bm{r'})$ is taken to be the Perey-Buck form \cite{Perey:1962}
\begin{align}
W_N(\bm{r},\bm{r}')=V_N[(r+r')/2]\frac{\exp[-(\bm{r}-\bm{r}')^2/\beta^2]}{\pi^{3/2}\beta^3}.
\label{PB}
\end{align}
Here, $\beta=\beta_0/A_\text{red}$ denotes the range of nonlocality \cite{Jackson:1974xhu}, with $\beta_0=0.85$ fm being the range of nonlocality from neutron scattering \cite{Perey:1962} and $A_\text{red}$ being the reduced mass number in fusion problems. Expanding $W_N(\bm{r},\bm{r}')$ in partial waves, we have
\begin{align}
\!\!\!W^{(L)}_N(r,r')=&V_N[(r+r')/2]\nonumber\\
\times&4 rr'\,\frac{\exp[-({r}^2+{r'}^2)/\beta^2]}{\pi^{1/2}\beta^3} i_L\!\!\left(\frac{2}{\beta^2}rr'\right).
\label{WNL}
\end{align}
$i_L(z)$ is the modified spherical Bessel function of the first kind. 
The relative momentum $k_L(r)$ in Eq.~\eqref{IWBC1} is estimated by $k_L(r)\!\!=\!\!\sqrt{2\mu\!\left[E-\frac{L(L+1)}{2\mu r^2}-V_N(r)-V_C(r)\right]}$
and $k_L(r)\!=\!\!\sqrt{2\mu\!\left[E-\frac{L(L+1)}{2\mu r^2}-W^\text{LE}_N(r)-V_C(r)\right]}$
%\begin{align}
%\!\!k_L(r)\!=\!
%\begin{cases}
%\sqrt{2\mu\!\left[E-\frac{L(L+1)}{2\mu r^2}-V_N(r)-V_C(r)\right]},\\[2ex]
%\sqrt{2\mu\!\left[E-\frac{L(L+1)}{2\mu r^2}-W^\text{LE}_N(r)-V_C(r)\right]},
%\end{cases}
%\end{align}
for local and nonlocal potentials, respectively.
$W_N^\text{LE}(r)=\exp\left\{-\frac{\mu\beta^2}{2}\left[E-V_C(r)-W_N^\text{LE}(r)\right]\right\}V_N(r)$ is the local equivalence of the Perey-Buck potential $W_N(\bm{r},\bm{r}')$ \cite{Perey:1962}.
%\begin{align}
%W_N^\text{LE}(r)\exp\left\{\frac{\mu\beta^2}{2}\left[E-V_C(r)-W_N^\text{LE}(r)\right]\right\}=V_N(r).
%\end{align}
%
%
%
In numerical calculations, we take the absorption radius to be $r_\text{abs}=A_P^{1/3}+A_T^{1/3}$ for all the partial waves.
This is slightly different from the CCFULL's convention where $r_\text{abs}$ is taken to be the local minimum of the total two-body potential (including the centrifugal potential) inside the Coulomb barrier.
We have verified explicitly that these two choices give fusion cross sections numerically close to each other.
The maximal partial-wave angular momentum is taken to be $L_\text{max}=10$.
Such a truncation gives rise to the truncation errors in theoretical results.
It is important to distinguish them from numerical errors from solving the Bloch-Schr\"odinger equations.
Our numerical codes are written by using arbitrary-precision arithmetic. 
They inevitably become less efficient than numerical codes written in double precision.
But, they allow us to handle the so-called ``ill-conditioned'' matrices in a more straightforward way, without introducing extra nuggets by hand to regularizing these matrices \cite{Furnstahl:2020abp}.
It thus helps us better understand our new theoretical tools.

We first calculate the fusion cross sections of the ${}^{14}\text{N}+{}^{12}\text{C}$ fusion reaction by using the generalized calculable $R$-matrix theory on a Lagrange mesh.
We take 
\begin{align}
V_0\!=\!60 \text{ MeV},\ \ R_0\!=\!1.206(A_T^{1/3}\!\!+\!A_P^{1/3}) \text{ fm},%\nonumber\\
\ \ a\!=\!0.5 \text{ fm},
\label{WSP}
\end{align}
for both the local and nonlocal potentials.
The channel radius is taken to be $r_\text{c}=20$ fm unless otherwise mentioned.
%The number of Lagrange functions is taken to be $N\geq20$ in numerical calculations.
%Fusion cross sections of the ${}^{14}\text{N}+{}^{12}\text{C}$ fusion reaction are plotted in Fig.~\ref{RM}, along with numerical relative errors.
%The theoretical results are plotted in Fig.~\ref{RM}, along with the relative errors.
In Fig.~\ref{RM}(a), the fusion cross sections are presented for both the local and nonlocal potentials, along with experimental data from Ref.~\cite{Switkowski:1977wsf}.
For the local and nonlocal potentials, we take the numbers of Lagrange functions to  be $N\geq20$ and $N\geq100$, respectively.
One can see that the theoretical results from the local and nonlocal potentials are almost indistinguishable in practice.
This could be inferred from the tiny range of nonlocality $\beta\sim0.13$ fm used in the nonlocal potential,
which suppresses nonlocality significantly and makes the nonlocal potential nearly diagonal in $r$-$r'$ space.
Such a ``singular'' nonlocal potential might be less interesting from the physical viewpoint.
But, it turns out to be a technical challenge for numerical calculations and provides an ideal playground to test the robustness of the generalized calculable $R$-matrix theory, as well as our numerical implementations in arbitrary precision.
%A large number of mesh points are needed in matrix-element computations to achieve good numerical precision.
It is found that more than 100 Lagrange functions are needed for numerical convergence in the case of the nonlocal potential, 
and our arbitrary-precision codes give the reliable numerical results. 
We also test our numerical codes for mild nonlocal potentials with larger ranges of nonlocality.
It is found that much fewer Lagrange functions are needed for moderate precision goals in these cases.
In Figs.~\ref{RM}(b) and \ref{RM}(c), we calculate $\text{relative}\ \text{errors}=(\text{theoretical}\ \text{results}-\text{benchmark}\ \text{results})/\text{benchmark results}$ of the generalized calculable $R$-matrix theory for different numbers of Lagrange functions. 
%For benchmark fusion cross sections, we take theoretical results with $N=200$ Lagrange functions.
For the local potential, it is found that the relative errors of the $N=20, 50, 100$ results are around $0.01$, $10^{-10}$, and $10^{-25}$, respectively, with the benchmark results taken at $N=200$.
Similarly, for the nonlocal potential, it is found that the relative errors of the $N=100,200,300$ results are around $0.01$, $10^{-8}$, and $10^{-15}$, respectively, with the benchmark results taken at $N=400$. 
These results show that the generalized calculable $R$-matrix theory generally gets convergent quickly with respect to the increasing numbers of Lagrange functions.
We would like to emphasize that the relative errors here are the numerical errors from solving the Bloch-Schr\"odinger equations and do not include the truncation errors from the partial-wave truncations.
The latter are found to be around $10^{-8}$-$10^{-3}$ for $E\in[3, 10]$ MeV. 

We then explore the parameter space of the ${}^{14}\text{N}+{}^{12}\text{C}$ fusion reaction with the generalized EC.
EC and its generalizations have been shown to be useful for quantifying theoretical uncertainties and analyzing global sensitivity in bound-state and scattering problems.
Our generalized EC aims to extend the method to low-energy heavy-ion fusion reactions.
The local potentials in the Woods-Saxon form are adopted to validate the generalized EC.
Numerical calculations with the nonlocal potentials in the Perey-Buck form are similar 
but more time-consuming. 
We treat $V_0$ as the free parameter and take its value between 40 MeV and 240 MeV. 
%We check the numerical precision of generalized EC in the ${}^{14}\text{N}+{}^{12}\text{C}$ fusion reaction over $E\in[3,10]$ MeV. 
Two sets of training points are examined, i.e., Training Set (a) with six training points at $V_0=45, 65, \cdots, 125, 145$ MeV and Training Set (b) with ten training points at $V_0=45, 55, \cdots, 125, 135$ MeV.
In Fig.~\ref{EC}, we plot the absolute values of the relative errors of the fusion cross sections given by the generalized EC at the test points $V_0=40, 60, \cdots, 220, 240$ MeV.
In Fig.~\ref{EC}(a), the relative errors of the fusion cross sections are found to be $\sim10^{-3}$-$0.4$ for Training Set (a) and depend strongly on the test points.
In general, relative errors at the interpolating points $V_0=60, \cdots, 100, 120$ MeV are much smaller than those at the extrapolating points $V_0=40, 160, \cdots, 220, 240$ MeV.
In other words, the generalized EC shows better performance for interpolation than extrapolation.
In Fig.~\ref{EC}(b), we increase the EC basis size from $N_\text{EC}=6$ to $N_\text{EC}=10$.
The relative errors get decreased systematically for all the test points and are found to be around $10^{-10}$ to $10^{-2}$.
Once again, the generalized EC gives better results for the interpolating points.

We also calculate the fusion cross sections at different channel radii.
The local potential in Eqs.~\eqref{WS} and \eqref{WSP} is used.
For concreteness, we take the reaction energy $E=5$ MeV and 
the channel radius $r_\text{c}$ between 20 fm and 1000 fm.
%For the ${}^{14}\text{N}+{}^{12}\text{C}$ fusion reaction considered here, nuclear interactions could be ignored safely at $r_\text{c}>20$ fm already.
A channel radius as large as 1000 fm is certainly not necessary for the ${}^{14}\text{N}+{}^{12}\text{C}$ fusion reaction.
But it gives a valuable chance to explore the advantages and limitations of our generalizations of the calculable $R$-matrix theory and EC.
Also,
there are important cases where channel radii as large as $\sim1000$ fm are indispensable to get physically meaningful results (e.g., Refs.~\cite{Descouvemont:2005rc,Nguyen:2011aa,Nguyen:2012uz,Thompson:2000ny}), 
and our experience with large channel radii could be helpful.
The numerical results are given in Fig.~\ref{rmax}.
%Generally, the consistency of calculable $R$-matrix theory requires that fusion cross sections should not depend on channel radii.
In Fig.~\ref{rmax}(a), the fusion cross sections are given by the generalized calculable $R$-matrix theory with $N=20, 50, 100, 200, 400, 800$ Lagrange functions at the channel radii $r_\text{c}\in[20, 1000]$ fm.
It is found that, the larger the channel radius is, the more Lagrange functions are needed to achieve numerical convergence.
For example, 20, 100, 800 Lagrange functions are needed for the channel radii $r_\text{c}=20, 100, 1000$ fm, respectively. 
In Fig.~\ref{rmax}(b), the fusion cross sections are given by the generalized EC 
from
the same $N_\text{EC}=6, 10$ training points as Fig.~\ref{EC}. 
The EC basis functions $\{\chi_L^\text{int}(\bm{\alpha}_i^\text{tr})\}$ are taken to be the exact solutions of the Bloch-Schr\"odinger equations at each channel radius, and the relative errors of fusion cross sections from the generalized EC are found to be about $2\times10^{-4}$ and $2\times10^{-9}$ for $N_\text{EC}=6$ and 10, respectively, which remain stable over the whole channel-radius interval $[20,1000]$ fm. 
Noticeably, to achieve similar numerical precision at large channel radii, the total number of basis functions needed in the generalized EC is significantly smaller than that in the generalized calculable $R$-matrix theory.
This contributes to the higher numerical efficiency of the generalized EC in scanning the parameter space than the generalized calculable $R$-matrix theory, especially at large channel radii. 
Also, it is interesting to note that in our cases the relations between the internal wave functions and the EC basis functions (i.e., Eq.~\eqref{GEC}) learned at the small channel radii remain valid approximately at the large channel radii.
Explicit calculations show that $\{c_i\}$ at different channel radii are indeed numerically close to each other.
In other words, the generalized EC works out universal relations among the internal wave functions at different points in the parameter space.
These relations remain valid even when the channel radius is enlarged from 20 fm to 1000 fm and the internal region is inflated by a factor of more than 50 along the radial direction.
%This could be verified by computing $\{c_i\}$ at different channel radii. 
This is drastically different from the generalized $R$-matrix theory $+$ the Lagrange functions, where the relations between the internal wave functions and the Lagrange functions are by no means universal and change significantly at different channel radii.

%Last, we would like to comment on numerical efficiency of our generalizations of calculable $R$-matrix theory and EC.
 
\section{Conclusions}

\label{Concl}

%\vskip -1cm

IWBC assumes incoming-wave profiles near the coordinate origins.
It is widely used in theoretical studies of low-energy heavy-ion fusion reactions to simulate the strong absorption inside the Coulomb barriers.
In this work, we generalize the calculable $R$-matrix theory and EC to IWBC.
The calculable $R$-matrix theory has been formulated for regular boundary conditions and applied successfully to study various structural and reaction problems in nuclear physics.
On the other hand, EC is a variational emulator to calculate physical observables at different points in the parameter space.
It has been worked out for bound-state and scattering observables.
%In this work, we generalize calculable $R$-matrix theory to IWBC, which is widely used in theoretical studies of low-energy heavy-ion fusion reactions to simulate strong absorption inside Coulomb barriers.
In the generalized calculable $R$-matrix theory, we divide configuration space into three regions by the absorption radius $r_\text{abs}$ and the channel radius $r_\text{c}$ and solve the
%Wave functions in different regions are matched continuously with each other by Bloch operators. 
 Bloch-Schr\"odinger equations for internal wave functions on the Lagrange meshes
to extract  fusion observables.
The generalized $R$-matrix theory then provides a natural starting point to generalize EC to fusion observables.
%which emulates fusion observables at different points in the parameter space by starting with a small set of training data.
% It could be useful in problems like uncertainty quantification and global sensitivity analysis.
As a proof of concept, we use our generalizations of the calculable $R$-matrix theory and EC to study the ${}^{14}\text{N}+{}^{12}\text{C}$ fusion reaction.
 Both local and non-local potentials are used in the numerical calculations.
%Both local and nonlocal potentials are used to describe nuclear interactions between ${}^{14}$N and ${}^{12}$C.
We check the numerical reliability of our generalizations of the calculable $R$-matrix theory and EC systematically.
 Both of them are found to work well for IWBC.

\begin{acknowledgments}

%\section*{Acknowledgments}

%\vskip -1cm

%D.~B.~would like to thank Dillon Frame for communications.
%This work is supported by the National Natural Science Foundation of China (Grants No.\ 12035011, No.\ 11905103, No.\ 11947211, No.\ 11535004, No.\ 11975167, No.\ 11761161001, No.\ 11565010, No.\ 11961141003), by the National Key R\&D Program of China (Contracts No.\ 2018YFA0404403 and No.\ 2016YFE0129300), by the Science and Technology Development Fund of Macau under Grant No.~008/2017/AFJ, by the Fundamental Research Funds for the Central Universities (Grant No.\ 22120200101), and by the China Postdoctoral Science Foundation (Grants No.\ 2019M660095 and No.\ 2020T130478).

This work is supported by the National Natural Science Foundation of China (Grants No.\ 12035011, No.\ 11905103, No.\ 11947211, No.\ 11535004, No.\ 11975167, No.\ 11761161001, No.\ 11565010, No.\ 11961141003, and No.\ 12022517), by the National Key R\&D Program of China (Contracts No.\ 2018YFA0404403 and No.\ 2016YFE0129300), by the Science and Technology Development Fund of Macau (Grants No.\ 0048/2020/A1 and No.\ 008/2017/AFJ), by the Fundamental Research Funds for the Central Universities (Grant No.\ 22120200101), and by the China Postdoctoral Science Foundation (Grants No.\ 2020T130478 and No.\ 2019M660095).

\end{acknowledgments}


\begin{widetext}

\appendix

\section{Lagrange Functions}
\label{LF}

Let $\{x_i,\lambda_i\}$ be the abscissae and weights associated with the Gauss quadrature on the interval $[0,1]$ \cite{Baye:2015,Press:1992},
\begin{align}
\mathbb{P}_N(2x_i-1)=0,\qquad \lambda_i=\frac{1}{4x_i(1-x_i)[\mathbb{P}_N'(2x_i-1)]^2},
\end{align}
with $i=1,\cdots,N$.
Here, $\mathbb{P}_N(x)$ is the Legendre polynomial of order $N$ and $\mathbb{P}_N'(x)\equiv \mathrm{d}\mathbb{P}_N(x)/\mathrm{d}x$.
The Lagrange functions $\{\mathbb{L}^{\!\!N}_i(r)\}$ are given by
\begin{align}
\mathbb{L}^{\!\!N}_i(r)=(-1)^{N+i}\sqrt{x_i(1-x_i)\Delta r}\frac{\mathbb{P}_N[(2r-r_\text{abs}-r_\text{c})/\Delta r]}{r-x_i\Delta r-r_\text{abs}},
\end{align}
with $\Delta r=r_\text{c}-r_\text{abs}$. It is straightforward to show that $\mathbb{L}^{\!\!N}_i(r_\text{abs}+x_j\Delta r)=(\Delta r\lambda_i)^{-1/2}\delta_{ij}$.
%\begin{align}
%\varphi_i(r_\text{abs}+x_j\Delta r)&=(-1)^{N+i}\sqrt{\frac{x_i(1-x_i)}{\Delta r}}\frac{P_N(2x_j-1)}{x_j-x_i}\nonumber\\
%&=(-1)^{N+i}\sqrt{\frac{x_i(1-x_i)}{\Delta r}}\frac{1}{x_j-x_i}\left[P_N(2x_i-1)+2P_N'(2x_i-1)(x_j-x_i)+\cdots\right]\nonumber\\
%&=(-1)^{N+i}\sqrt{\frac{x_i(1-x_i)}{\Delta r}}2P_N'(2x_i-1)\nonumber\\
%&=(\Delta r\lambda_i)^{-1/2}\delta_{ij},
%\end{align}
%where $P'_N(2x_i-1)=(-1)^{N+i}[4x_i(1-x_i)\lambda_i]^{-1/2}$ is used in the simplification.
 The relevant matrix elements of the Lagrange functions are given as follows:
\begin{align}
\left(\mathbb{L}^{\!\!N}_i|\mathbb{L}^{\!\!N}_j\right)&\equiv\int_{r_\text{abs}}^{r_\text{c}}\!\mathrm{d}r\,\mathbb{L}^{\!\!N}_i(r)\,\mathbb{L}^{\!\!N}_j(r)
%\approx\Delta r\sum_{k=1}^N\lambda_k\varphi_i(r_\text{abs}+x_k\Delta r)\varphi_j(r_\text{abs}+x_k\Delta r)\nonumber\\
%&=\Delta r\sum_{k=1}^N\lambda_k(\Delta r\lambda_i)^{-1/2}\delta_{ik}(\Delta r\lambda_j)^{-1/2}\delta_{jk}\nonumber\\
=\delta_{ij},\\
%\end{align}
%
%\begin{align}
\left(\mathbb{L}^{\!\!N}_i|V(r)|\mathbb{L}^{\!\!N}_j\right)&\equiv\int_{r_\text{abs}}^{r_\text{c}}\!\mathrm{d}r\,\mathbb{L}^{\!\!N}_i(r)V(r)\,\mathbb{L}^{\!\!N}_j(r)%\nonumber\\
%&\approx\Delta r\sum_{k=1}^N\lambda_k\varphi_i(r_\text{abs}+x_k\Delta r)V(r_\text{abs}+x_k\Delta r)\varphi_j(r_\text{abs}+x_k\Delta r)\nonumber\\
%&=\Delta r\sum_{k=1}^N\lambda_k(\Delta r\lambda_i)^{-1/2}\delta_{ik}V(r_\text{abs}+x_k\Delta r)(\Delta r\lambda_j)^{-1/2}\delta_{jk}\nonumber\\
=\delta_{ij}V(r_\text{abs}+x_i\Delta r),\\
\left(\mathbb{L}^{\!\!N}_i|W(r,r')|\mathbb{L}^{\!\!N}_j\right)&\equiv\int_{r_\text{abs}}^{r_\text{c}}\!\mathrm{d}r\!\int_{r_\text{abs}}^{r_\text{c}}\!\mathrm{d}r'\,\mathbb{L}^{\!\!N}_i(r)W(r,r')\,\mathbb{L}^{\!\!N}_j(r')\nonumber\\
%&\approx(\Delta r)^2\sum_{m=1}^N\sum_{n=1}^N\lambda_m\lambda_n\varphi_i(r_\text{abs}+x_m\Delta r)W(r_\text{abs}+x_m\Delta r,r_\text{abs}+x_n\Delta r)\varphi_j(r_\text{abs}+x_n\Delta r)\nonumber\\
%&=(\Delta r)^2\sum_{m=1}^N\sum_{n=1}^N\lambda_m\lambda_n(\Delta r\lambda_i)^{-1/2}\delta_{im}W(r_\text{abs}+x_m\Delta r,r_\text{abs}+x_n\Delta r)(\Delta r\lambda_j)^{-1/2}\delta_{jn}\nonumber\\
&
=\Delta r\sqrt{\lambda_i\lambda_j}\,W(r_\text{abs}+x_i\Delta r,r_\text{abs}+x_j\Delta r),
\end{align}
\begin{align}
\left(\mathbb{L}^{\!\!N}_i\left|\frac{\mathrm{d}^2}{\mathrm{d}r^2}\right|\mathbb{L}^{\!\!N}_j\right)&\equiv\int_{r_\text{abs}}^{r_\text{c}}\!\mathrm{d}r\,\mathbb{L}^{\!\!N}_i(r)\,{\mathbb{L}^{\!\!N}_j}''(r)\nonumber\\
%\approx\Delta r\sum_{k=1}^N\lambda_k\varphi_i(r_\text{abs}+x_k\Delta r)\varphi_j''(r_\text{abs}+x_k\Delta r)\nonumber\\
%=(\Delta r\lambda_i)^{1/2}\varphi''_j(r_\text{abs}+x_i\Delta r).
&=
\begin{cases}
\frac{\left(N^2+N+6\right) \left(x_i-1\right) x_i+2}{3 {\Delta r}^2
   \left(x_i-1\right){}^2 x_i^2}, & i=j,\\[2ex]
 \frac{(-1)^{i+j} \sqrt{\left(x_i-1\right) \left(x_j-1\right) x_j} \left[x_i \left(4
   x_i-2 x_j-3\right)+x_j\right]}{{\Delta r}^2 \left(x_i-1\right){}^2 x_i^{3/2}
   \left(x_i-x_j\right){}^2},  & i\neq j,
\end{cases}
\\
%\end{align}
%For $i=j$, we have
%\begin{align}
%\left(\varphi_i\left|\frac{\mathrm{d}^2}{\mathrm{d}r^2}\right|\varphi_i\right)=\frac{\left(N^2+N+6\right) \left(x_i-1\right) x_i+2}{3 {\Delta r}^2
%   \left(x_i-1\right){}^2 x_i^2}.
%\end{align}
%while for $i\neq j$,
%\begin{align}
%\left(\varphi_i\left|\frac{\mathrm{d}^2}{\mathrm{d}r^2}\right|\varphi_j\right)=\frac{(-1)^{i+j} \sqrt{\left(x_i-1\right) \left(x_j-1\right) x_j} \left[x_i \left(4
%   x_i-2 x_j-3\right)+x_j\right]}{{\Delta r}^2 \left(x_i-1\right){}^2 x_i^{3/2}
%   \left(x_i-x_j\right){}^2}.
%\end{align}
%
%\begin{align}
\left(\mathbb{L}^{\!\!N}_i\left|\delta(r-r_\text{abs})\frac{\mathrm{d}}{\mathrm{d}r}\right|\mathbb{L}^{\!\!N}_j\right)&\equiv\int_{r_\text{abs}}^{r_\text{c}}\!\mathrm{d}r\,\mathbb{L}^{\!\!N}_i(r)\,{\mathbb{L}^{\!\!N}_j}'(r)\,\delta(r-r_\text{abs})
%=\varphi_i(r_\text{abs})\varphi_j'(r_\text{abs})
\nonumber\\
&=\frac{(-1)^{i+j+1} \sqrt{\left(x_i-1\right) x_i \left(x_j-1\right) x_j} \left[N (N+1)
   x_j-1\right]}{{\Delta r}^2 x_i x_j^2},\\
%\end{align}
%
%\begin{align}
\left(\mathbb{L}^{\!\!N}_i\left|\delta(r-r_\text{c})\frac{\mathrm{d}}{\mathrm{d}r}\right|\mathbb{L}^{\!\!N}_j\right)&\equiv\int_{r_\text{abs}}^{r_\text{c}}\!\mathrm{d}r\,\mathbb{L}^{\!\!N}_i(r)\,{\mathbb{L}^{\!\!N}_j}'(r)\,\delta(r-r_\text{c})
%=\varphi_i(r_\text{c})\varphi_j'(r_\text{c})
\nonumber\\
&=\frac{(-1)^{i+j+1} x_i 
   \left[N (N+1) \left(x_j-1\right)+1\right]}{{\Delta r}^2
   \left(x_j-1\right){}^2}\sqrt{\frac{\left(x_j-1\right) x_j}{\left(x_i-1\right) x_i}}.
\end{align}

%\section{Nonlocal Potentials}
%
%\begin{align}
%W_N(\bm{r},\bm{r}')=V_N[(r+r')/2]\frac{\exp[-(\bm{r}-\bm{r}')^2/\beta^2]}{\pi^{3/2}\beta^3}.
%\end{align}
%
%\begin{align}
%&\int\!\mathrm{d}^3r'\,W_N(\bm{r},\bm{r}')\,\Psi(\bm{r}')\nonumber\\
%=&\int\!\mathrm{d}^3r'\,W_N(\bm{r},\bm{r}')\,\sum_{LM}\frac{\chi_L(r')}{r'}Y_{LM}(\widehat{r'})\nonumber\\
%=&\sum_{LM}\int\!\mathrm{d}^3r'\,W_N(\bm{r},\bm{r}')\,\frac{\chi_L(r')}{r'}Y_{LM}(\widehat{r'})\nonumber\\
%=&\sum_{LM}\int\!\mathrm{d}r'\,r'^2\int\!\mathrm{d}^2\widehat{r'}\,V_N[(r+r')/2]\frac{\exp[-(\bm{r}-\bm{r}')^2/\beta^2]}{\pi^{3/2}\beta^3}\frac{\chi_L(r')}{r'}Y_{LM}(\widehat{r'})\nonumber\\
%=&\sum_{LM}\int\!\mathrm{d}r'\,r'\,V_N[(r+r')/2]\frac{\exp[-({r}^2+{r'}^2)/\beta^2]}{\pi^{3/2}\beta^3}{\chi_L(r')}\int\!\mathrm{d}^2\widehat{r'}\,\exp\!\left(\frac{2}{\beta^2}\bm{r}\!\cdot\!\bm{r}'\right)Y_{LM}(\widehat{r'})\nonumber\\
%=&4\pi\sum_{LM}\sum_{L'M'}\int\!\mathrm{d}r'\,r'\,V_N[(r+r')/2]\frac{\exp[-({r}^2+{r'}^2)/\beta^2]}{\pi^{3/2}\beta^3}{\chi_L(r')}\nonumber\\
%\times&\int\!\mathrm{d}^2\widehat{r'}\,(-1)^{L'}i_{L'}\!\!\left(-\frac{2}{\beta^2}rr'\right)Y_{L'M'}(\widehat{r})\,Y_{L'M'}(\widehat{r'})^*Y_{LM}(\widehat{r'})\nonumber\\
%=&4\pi\sum_{LM}\int\!\mathrm{d}r'\,r'\,V_N[(r+r')/2]\frac{\exp[-({r}^2+{r'}^2)/\beta^2]}{\pi^{3/2}\beta^3} i_L\!\!\left(\frac{2}{\beta^2}rr'\right){\chi_L(r')}Y_{LM}(\widehat{r}).
%\end{align}
%
%\begin{align}
%W^{(L)}_N(r,r')=4\pi rr'\,V_N[(r+r')/2]\,\frac{\exp[-({r}^2+{r'}^2)/\beta^2]}{\pi^{3/2}\beta^3} i_L\!\!\left(\frac{2}{\beta^2}rr'\right).
%\end{align}
%
%\begin{align}
%W^\text{LE}_N(r)
%\end{align}
%
\end{widetext}
%
%In the above derivation, 
%the Gauss quadrature formula for the integral of an arbitrary function $f(r)$ over $[r_\text{abs},r_\text{c}]$ is used
%\begin{align}
%\int_{r_\text{abs}}^{r_\text{c}}\!\mathrm{d}r\,f(r)\approx{\Delta r}\sum_{i=1}^N\lambda_i\,f\!\left({r_\text{abs}}+x_i{\Delta r}\right).
%\end{align}
%\begin{align}
%\int_0^1\!\mathrm{d}x\,f(x)\approx\sum_{i=1}^N\lambda_i\,f(x_i).
%\end{align}
%Therefore, all the matrix elements of the Bloch-Schr\"odinger equation can be calculated analytically.
%Compared with other choices of variational bases, this is an important advantage of Lagrange functions and gives rise to efficient and stable numerical codes.

\end{document}